\documentclass[12pt,preprint]{aastex}

\newcommand   {\mic}   {\mbox{$\mu$m}}

\renewcommand {\deg}   {\mbox{$^\circ$}}

\newcommand   {\kms}   {\mbox{km\,s$^{-1}$}}
\renewcommand {\ga}    {\mbox{\rlap{\hbox{\lower5pt\hbox{$\sim$}}}\hbox{$>$}}}
\renewcommand {\la}    {\mbox{\rlap{\hbox{\lower5pt\hbox{$\sim$}}}\hbox{$<$}}}

\newcommand {\sgra}  {\mbox{Sgr\ A$^*$}}

\def\ee #1 {\times 10^{#1}}
\def\ut #1 #2 { \, \mathrm{#1}^{#2}}
\def\u #1 { \, \mathrm{#1}}
\def\kms {\,\mathrm{km\,s}^{-1}}

\def\la{\lower.4ex\hbox{$\;\buildrel <\over{\scriptstyle\sim}\;$}}
\def\ga{\lower.4ex\hbox{$\;\buildrel >\over{\scriptstyle\sim}\;$}}

\begin{document}



\def\kms {\hbox{km{\hskip0.1em}s$^{-1}$}} 
\def\msol{\hbox{$\hbox{M}_\odot$}}
\def\lsol{\hbox{$\hbox{L}_\odot$}}
\def\kms{km s$^{-1}$}
\def\Blos{B$_{\rm los}$}
\def\etal   {{\it et al. }}                     
\def\psec           {$.\negthinspace^{s}$}
\def\pasec          {$.\negthinspace^{\prime\prime}$}
\def\pdeg           {$.\kern-.25em ^{^\circ}$}
\def\degree{\ifmmode{^\circ} \else{$^\circ$}\fi}
\def\ee #1 {\times 10^{#1}}          
\def\ut #1 #2 { \, \textrm{#1}^{#2}} 
\def\u #1 { \, \textrm{#1}}          
\def\nH {n_\mathrm{H}}

\def\ddeg   {\hbox{$.\!\!^\circ$}}              
\def\deg    {$^{\circ}$}                        
\def\le     {$\leq$}                            
\def\sec    {$^{\rm s}$}                        
\def\msol   {\hbox{$M_\odot$}}                  
\def\i      {\hbox{\it I}}                      
\def\v      {\hbox{\it V}}                      
\def\dasec  {\hbox{$.\!\!^{\prime\prime}$}}     
\def\asec   {$^{\prime\prime}$}                 
\def\dasec  {\hbox{$.\!\!^{\prime\prime}$}}     
\def\dsec   {\hbox{$.\!\!^{\rm s}$}}            
\def\min    {$^{\rm m}$}                        
\def\hour   {$^{\rm h}$}                        
\def\amin   {$^{\prime}$}                       
\def\lsol{\, \hbox{$\hbox{L}_\odot$}}
\def\sec    {$^{\rm s}$}                        
\def\etal   {{\it et al. }}                     

\def\xbar   {\hbox{$\overline{\rm x}$}}         


\title{HST Observations of the Stellar Distribution Near \sgra}
\author{F. Yusef-Zadeh$^1$, H. Bushouse$^2$ \& M. Wardle$^{3,4}$}
\affil{$^1$Department of Physics and Astronomy, Northwestern University, Evanston, IL 60208}
\affil{$^2$STScI, 3700 San Martin Drive, Baltimore, MD  21218}
\affil{$^3$Macquarie University Research Centre in Astronomy, Astrophysics 
and Astrophotonics} 
\affil{$^4$Department of Physics and Astronomy, Macquarie University, 
Sydney NSW 2109, Australia}

\begin{abstract}

\sgra\ is embedded within the nuclear cluster, which consists of a mixture of evolved and young 
populations of stars dominating the light over a wide range of angular scales. Here we present 
HST/NICMOS data to study the surface brightness distribution of stellar 
light  within the inner 10$''$ of 
\sgra\ at 1.45$\mu$m, 1.7$\mu$m and 1.9$\mu$m. We use these data to independently examine the 
surface brightness distribution that had been measured previously with NICMOS and  to determine 
whether there is a drop in the surface density of stars very near \sgra. Our analysis confirms 
that a previously reported drop in the surface brightness within 0.8$''$ of \sgra\ is 
an artifact of bright and massive stars near that radius.  We also show that the surface brightness
profile  within 5$''$ or $\sim$0.2 pc of \sgra\ can be fitted with broken power laws. 
The
 power laws are consistent with previous measurements, in that the profile becomes 
shallower at small radii. 
For radii $>$ 0.7\asec, the slope 
is $\beta=-0.34\pm0.04$ where $\Sigma$ is $\propto r^{\beta}$
and becomes flatter at smaller radii with $\beta=-0.13\pm0.04$.
Modeling of the surface brightness profile gives a stellar density that increases 
roughly as r$^{-1}$ within the inner 1$''$ of \sgra.  This slope confirms earlier measurements 
in that it is not consistent with that expected from an old, dynamically-relaxed stellar cluster
with a central supermassive black hole.  Assuming that the diffuse emission 
is not contaminated by a faint 
population of young stars down to the 17.1 magnitude limit of our imaging data at 1.70$\mu$, 
the shallow cusp profile is not 
consistent 
with a decline in stellar density in the inner arcsecond. 
In addition, 
converting our measured diffuse light profile to a stellar mass profile, with the assumption that the light is 
dominated by 
K0 dwarfs, the enclosed stellar mass 
within radius $r\la 0.1$\,pc of Sgr A* is $\approx 3.2\ee 4 $\,M$_\odot (r/0.1\,\mathrm{pc})^{2.1}$.  
\end{abstract}


\section{Introduction}
\label{introduction}

There is compelling evidence that the compact radio source \sgra\ is located at the very 
dynamical center of our galaxy (Reid \& Brunthaler 2004)
 and coincides with a $4 \times 10^6$\msol\ black 
hole  (Ghez et al. 2009; Gillessen et al. 2009).   The nuclear cluster surrounding \sgra\ consists of a mixture 
of an evolved stellar population, probably having an isotropic stellar light distribution (Trippe et al. 2008; 
Sch\"odel et 
al. 2009), and a young 
population of stars at smaller radii from \sgra .  \sgra\ is known to be less energetic and 
massive than black holes in luminous active galactic nuclei (AGN), which can outshine the brightness of 
their nuclear clusters.  Given the low luminosity of \sgra\ and its proximity, high spatial 
resolution observations of the nuclear cluster provide an unparalleled  opportunity to observe 
in detail the influence of the central black hole on the spatial distribution of stars and to 
test theories of cusp formation.  

Peebles (1972) showed
that the signature of a central black hole in a dense stellar cluster is a cusp 
with a power law form of $\rho\propto r^{\gamma}$.  Follow-up studies indicate that 
the power law slopes can range from $\gamma$=--3/2 to --7/4 depending on whether the mass of stars 
dominating the stellar density profile is  neutron stars, white dwarfs, or stellar black holes,
respectively (Bahcall and Wolf 1976, 1977; see reviews by Merritt 2006, Alexander 2005; 
Genzel et al. 2010).  Numerical simulations suggest  that 
if the stellar cluster is dynamically relaxed, it  should have a 
cusp with a power law slope in density of $-$2 for massive stars and and $-$1.5 for low mass 
stars (Hopman \& Alexander 2006; Preto \& Amaro-Seoane 2010). 


Due to its proximity, the Galactic center is an ideal place to examine the influence of 
\sgra\ on the spatial distribution of evolved stars. Late-type K and M giant 
stars could be used as good tracers of cusp dynamics, assuming they are older than the 
relaxation time scale (Alexander et al. 2007; Sch\"odel et al. 2007; Merritt 2010).  
Numerous measurements have been carried out 
over 
the years to study the infrared distribution of stars in the vicinity of \sgra\ and 
investigate the power law distribution of evolved stars in the hope of finding evidence for
a stellar cusp. 
The study by  Becklin \&  Neugebauer 
(1968) revealed  that on parsec size scales 
the surface brightness ($\Sigma$) profile 
is a power 
law. Numerous follow-up studies confirmed that 
 the slope beyond the projected distance r=0.5pc follows a power 
law slope of $\beta$=--0.8, where 
$\Sigma$  $\propto r^{\beta}$ 
($\gamma$=$\beta$+1)
(e.g., Catchpole, Whitelock, \& Glass 1990; McGinn et al. 
1989; Rieke and Rieke 1994; Haller et al. 1996). 
The signature of a stellar cusp, however, should be seen within the inner 0.5pc, it
has proved more challenging to measure the slope because of source crowding,  the presence 
of bright stars and possible differential extinction. 
Early studies indicated that stellar 
number density measurements show a flat stellar cluster within a 
projected radius of 0.2--0.3pc (McGinn et al. 1968; Eckart et a. 1993; Haller et al. 1996; 
Launhardt et al. 2002). Follow-up studies used Adaptive Optics (AO) to improve the 
sensitivity and spatial resolutions and determined that the volume density of stars within 
$\sim$0.25--0.35 pc increases toward \sgra\ with $\gamma\sim$-1.3 to -1.4 (Genzel et al. 2003).
 Using better quality data and  improved methods of analysis, 
Sch\"odel et al. (2007)
estimated the power law slope  $\gamma=-1.2\pm0.05$
at a projected radius of 6$''$. 
This slope is not consistent with that expected from an old, dynamically-relaxed stellar 
cluster with a central supermassive black hole.

Unlike the technique of measuring stellar density counts of stars, Scoville et al. (2003)
used HST/NICMOS images to measure the diffuse stellar emission near \sgra, 
determining the radial distribution of observed and extinction-corrected 1.9$\mu$m emission. 
The spatial variation of extinction was determined by using 
Pa$\alpha$, radio continuum and radio recombination line measurements. The greatest 
difference between the observed and extinction-corrected distribution occurs at projected 
radii $>$10\asec\ (corresponding to deprojected radii $\sim$30\asec), where the larger extinction 
corrections associated with the circumnuclear molecular ring raises the extinction-corrected 
flux by a greater amount. They showed that the observed and extinction-corrected 1.9$\mu$m
surface brightness distribution increases strongly inwards to a projected radius of 
1\asec\ and then exhibits a drop within 0.8\asec\ of \sgra. The drop in the surface 
brightness distribution is an important parameter to determine accurately. This feature
could have several different origins: it might be the core radius of the nuclear stellar 
distribution, it might be due to the depletion of late-type stars with high mass-to-light 
ratios (e.g., Bailey \&\ Davies 1999), or due to depletion of giant stars through 
collisions with main-sequence stars (Genzel et al. 1996; Sellgren et al. 1990), or even 
ejection of stars from the central 1\asec\ by a past or present binary black hole (Hansen \& 
Milosavljevic 2003; Yu \& Tremaine 2003). 

The ideas explaining the depletion of late-type stars remained 
viable, as described below, but  the drop in the stellar surface brightness 
was shown to  be due to artifacts of bright, massive stars that had not been 
subtracted properly.  Sch\"odel et al. (2007) 
suspected that light density in the HST images is dominated
by the brightest stars and the wings of the PSF from the brightest
 stars. They created  artificial images from the
stellar number counts, to which they applied the method of Scoville et  al. (2003). 
The comparison led Sch\"odel et al. (2007) to conclude that the excess light 
density found by Scoville et al. (2003) at a projected distance of 1--5\asec, corresponding 
to projected radii of 0.039 and 0.4 pc at the 8 kpc distance to the Galactic center, was 
an artifact due to contamination of bright, young stars. These bright stars are now known to be associated 
with a clockwise rotating thick stellar disk, with surface density scaling as the inverse square 
of the true distance from \sgra\ (Genzel et al.\ 2003; Paumard et al.\ 2006; Bartko et al. 2010), 
making the true surface brightness distribution of the evolved cluster difficult to measure. 
The projected inner radius of the clockwise disk is $\sim$0.3 pc or $\sim$8\asec\ from \sgra, 
and the stellar ages are estimated to be $\sim$6 Myr, with the total mass of stars 
amounting to $\sim 1.5\times 10^4 \msol$ (Paumard et al.\ 2006).

The nuclear cluster at the Galactic center is unique in that the cluster can be resolved down to 
scales of milli-pc with AO measurements. However, the distribution of bright, 
young massive stars, which are not dynamically relaxed, creates difficulty in testing the cusp 
formation near \sgra. Apart from the fact that confusion limits the detection of stars fainter
than K$\sim$18--19,  an additional  factor that limits measurements of the radial profile of 
the  evolved cluster is identification of the young population of stars embedded within the 
nuclear cluster.  Earlier studies used broad-band imaging to determine the radial profile of 
stellar light near \sgra. Using imaging techniques, Sch\"odel et al. (2007) determined  a broken 
power-law distribution of stars with a break radius of $\sim$6\asec. At radii $r < 6$\asec, the 
surface number density profile $\Sigma$(r) is fitted by a power-law index $\beta=-0.19\pm0.05$, 
which is flatter than $\beta=-0.75\pm0.1$ at $r > 6$\asec. 
More recently, three separate studies have used narrow-band imaging (CO bandhead) or spectroscopy 
to separate the young and evolved stellar populations in the central 1\asec,
so that the radial profile of the evolved stellar system can be distinguished and measured 
accurately,  thus allowing theories of cusp formation to be tested. 
Strong CO bandhead absorption features in late type stars are used to distinguish them  
from early-type  stars.  Buchholz et al. (2009), following earlier studies by Genzel et al. (2003), 
used deep CO bandhead images (K $<$ 15.5) and found a shallow slope $\beta=0.17\pm0.09$. 
Genzel et al. (2003) had shown that the fraction of late-type stars decreases between 
distances of 10 and 1\asec\ from \sgra .  
In another study, Do et al. (2009) used the Br$\gamma$ and NaI  lines of stellar sources with 
K$'<$15.5 to distinguish between  early and late-type stars. 
They also found  a shallow slope of evolved stars with $\beta=0.27\pm0.19$.
Lastly, Bartko et al. (2010) used SINFONI to make spectroscopic identification of 
late-type and early-type stars using both CO bandheads and Br$\gamma$ lines. This study 
also showed a dearth of late-type stars at small distances from \sgra.
All of these results contradict earlier evidence for the signature of a cusp, and confirm 
a  depletion in the number density of stars close to \sgra\ (Genzel et al. 2003; 
Paumard et al. 2006; Sch\"odel et al. 2007; Buchholz et al. 2009; Do et al. 2009;  
Bartko et al. 2010). 
These measurements indicate that the stellar light distribution of evolved stars approaching \sgra\ 
becomes even shallower than that measured by Sch\"odel et al. (2007), implying a deficiency of 
evolved stars. 


A disadvantage of the NICMOS images is the relatively low 
spatial resolution, which can make contamination from the extended PSF wings of bright stars a problem. AO 
measurements have a clear advantage over HST images in that they can detect individual point sources. Another 
disadvantage of HST/NICMOS data is that it does not have spectroscopic capability to distinguish between late 
and early type stars.
However, measuring the surface brightness distribution using NICMOS images has advantages because of its 
sensitivity to 
low surface brightness diffuse light, the high Strehl ratio of the imaging, and PSF stability. 

The surface brightness distribution of extended and diffuse           
features are  measured as a function of the projected distance from SgrA*. 
This type of measurements can especially be useful 
within 1\asec\ (0.04pc) of \sgra, where
limited number of resolved stellar sources  are detected.

There are limited non-AO measurements of the surface brightness distribution 
of stars at the Galactic center that  have been attempted.  In particular, the distribution
of stars is not sampled well within 1\asec\ of \sgra, because there are so few stars
that can be identified spectroscopically within a limiting magnitude of K$\sim$15.5.
The functional form of the distribution of stars has therefore been uncertain.  
Here we study the surface brightness distribution at 1.45$\mu$m, 1.7$\mu$m and
1.9$\mu$m using NICMOS data originally obtained in September 2004 and April
2007 to monitor the near-IR flaring activity of \sgra.
Measurements presented here use a more extensive data set compared to that of
Scoville et al. (2003). 
In addition, a different technique has been used
to remove the bright and  massive stars from the surface brightness
measurements, as discussed in detail in the next section.  



\section{Data Reduction and Analysis}

For this study we have used NICMOS imaging obtained in September 2004 and April 2007,
as part of \sgra\ monitoring campaigns (HST GO programs 10179 and 10859, respectively).
All of the exposures used NICMOS camera 1, which
has a pixel scale of 0.043\asec\ and a field of view of $11'' \times 11''$. From the April 2007
data set we have used images obtained in the medium-band filters F145M and F170M, which have
central wavelengths of 1.46\mic\ and 1.71\mic, respectively, and a FWHM of 0.2\mic. From
the September 2004 data set, we have used images obtained in the narrow-band F190N filter,  
which has a central wavelength of 1.90\mic\ and a FWHM of 0.02\mic. There are 320 exposures
of 144 sec duration for each of the F145M and F170M filters, for a total exposure time of
768 minutes (12.8 hours) in each filter. For the F190N data, there are 64 exposures of
448 sec duration, for a total exposure time of 478 minutes (8 hours).

The sets of calibrated exposures for each filter were combined into a single image using
the STScI program ``multidrizzle''. Multidrizzle not only combines the aligned individual
exposures, but also rejects outliers (e.g. cosmic ray hits) and removes the geometric
distortions of the NICMOS optics. An example of the final F170M image is shown in
Figure 1a.

The radial distribution of surface brightness in the three combined images was
measured by performing aperture photometry using a routine written in Python, which  
duplicates the methods used in standard routines like the IRAF task ``phot''. The
custom Python routine allowed us to explore the results from various measurement
methods, as discussed below. Photometry was performed in a series of concentric circular    
apertures of increasing radius, centered on the position of \sgra.  The surface
brightness as a function of radius was then computed from the signal in the
annuli formed by each adjacent pair of apertures.

Our goal is to measure the surface brightness of the relaxed stellar
cluster surrounding Sgr A*, which is, unfortunately, contaminated by the
light from bright, young stars. Stars fainter than K~15.5-16 within the    
innermost arcseconds of Sgr A* are difficult to classify, nor do we have
the ability to resolve individual stars of this brightness in the HST
images. Hence the best we can do to remove the contribution of the young   
stars is to at least remove the light from the brightest ones that we can 
resolve. The remaining diffuse light could have contributions from stars
as young as types B to A, but we will assume for the sake of simplicity
that it is dominated by the more numerous old population.
There are several ways in which this could
be done. First, the bright, young stars could be removed from the combined
images in each filter band by PSF fitting and subtraction techniques. The
NICMOS camera 1 PSF, however, is quite complex and spatially-variant. Given
the extreme crowding in our images, it is virtually impossible to fit the PSF
to a sufficient level of accuracy to remove the more complex features, which
was verified by several different PSF subtraction attempts. Another way to
remove the contribution from bright stars is to simply not include their
signals when summing the data within the photometry apertures.  In the study
by Scoville et al. (2003), for example, which also used NICMOS images
to measure the surface brightness distribution around \sgra, photometry was
computed from the median pixel value after rejecting the brightest 20\%\ of the
pixels within each annulus. While this helps to remove some of the effect of
bright stars, it has the drawback of using a varying rejection threshold in each
annulus. Annuli containing very bright stars will have a much higher rejection
threshold and therefore still retain pixels associated with moderately bright
stars, resulting in inappropriate fluctuations in the radial light profile.
The sharp drop in surface brightness within 0.8\asec\ of \sgra\ found by
Scoville et al. (2003) was simply due to the presence of bright stars at
r$\sim$0.8\asec, which had not been completely removed by their median flux
technique (see also Sch\"odel et al. 2007).

The method we have adopted simply masks the pixels associated with bright stars
throughout the area of the images covered by our apertures before performing
the photometry measurements. The masking removes both the signal and the area
contributed by each pixel. NICMOS camera 1 oversamples the HST point spread 
function (PSF) at these wavelengths, resulting in stellar images that have a
core FWHM of nearly 4 pixels, and the first airy ring occurring at a radius of
$\sim$6 pixels.  We therefore used a mask with a radius of 9 pixels for the
brightest stars in the field, which completely blocks out the core and the  
first airy ring.  For fainter stars, where the signal from the first airy ring  
is negligible, we used a smaller mask of radius 3 pixels, which masks out only
the core of the PSF. The same set of stars was masked out in all three bands.
The faintest star masked out in the F170M image has an observed flux of
0.32 mJy in that band, which corresponds to a 1.7\mic\ observed magnitude of
16.2. Assuming $A_K = 2.42$ mag and
$A_{\lambda} \propto {\lambda}^{-2.11}$ (Fritz et al. 2011), these translate
to extinction-corrected values of 17.3 mJy and 11.9 mag.
We have not applied an extinction map to our data, because previously
derived maps have shown very little variation over the extent of our
relatively limited field of concern. A constant extinction value was
therefore adopted for each wavelength band.

We closely compared the HST image with available AO images (Figure 1 of
Gillessen et al. 2009) to identify and mask the brightest, young stars closest
to \sgra.  Figures 1a,b show a close-up view of this region of the
1.7\mic\ image with and without these stars masked.  Included in this masking
is the star S0-2, which is the closet star just to the north of \sgra\ and has
an observed 1.7\mic\ flux density of 0.75 mJy (15.3 mag).  We did not mask the
nearby stars S34/S35, because they are identified as evolved stars.

Evaluating the curve of growth of the NICMOS camera 1 PSF shows that the
9-pixel radius mask used for the brightest stars removes 89\%\ of the light of
a star, with the remaining 11\%\ in the unmasked features of the PSF wings.  
The brightest stars that have been masked in our images are $\sim$50--125 times
brighter than the faintest masked stars and therefore the unmasked 11\%\ of 
one of the bright stars would be equivalent to 6--14 additional faint stars
being included in the photometry.  Some of this extra signal is excluded
by the masking applied to nearby neighbors and the remaining amount is spread
across more than one radial bin. Given the amount of extra signal relative to
the total within a given annulus, the overall results presented here are not
affected in a qualitative way by this remaining signal.

The faintest diffuse emission remaining in the masked images has a
point-source equivalent observed 1.7\mic\ flux density of 0.14 mJy and, which
corresponds to an observed magnitude of 17.1.
We detect this diffuse emission within the central arcsecond with a
signal-to-noise ratio of $\sim$10. In the 1.45\mic\ and 1.9\mic\ images the
limiting values are 0.029 mJy and 19.1mag, and
0.32 mJy and 16.1mag, respectively. These values correspond to extinction
corrected point-source fluxes of 7.6--7.8 mJy and magnitudes of 12.6--13.0 
in all three bands. 

To measure the surface brightness, we used a series of photometric aperture 
sizes that
ranged from a radius of 1 pixel (0.043\asec) to 110 pixels (4.7\asec) around 
\sgra.  The largest aperture radius extends to the nearest edge of the images.
We chose increments in aperture radius ranging from 1 pixel (0.043\asec) near 
\sgra\ to 10 pixels (0.43\asec) near the edge of the image.
Figure 1 shows the series of photometry apertures overlaid on both the original
and masked versions of the F170M image.
Note that due to the very small number of unmasked pixels contained with the
smallest aperture (1 pixel radius) the data from that aperture were not used
in any of the subsequent analysis. Hence the smallest aperture used in the
remaining analysis corresponds a radius of two pixels (0.086\asec).

The resulting surface brightness distributions are shown for all three bands in Figure 2. These measurements include 
uncertainties due to Poisson noise in the sources, as well as uncertainty in the sky background. The plots in the right panel of 
Figure 2, which are extinction corrected, show that each band has about the same corrected flux (e.g. 0.4-0.5 Jy arcsec$^2$ at 
the 
smallest radius) suggesting that at least the wavelength dependence of the correction has been done correctly. We also evaluated 
if there is an indication of residual corrections in the spatial dimension. 
We studied the behavior of the three curves of 
Figure 2 as a function of radius. We overlayed the 3 curves on top of one another and noted the three 
plots run parallel to one 
another as a function of radius. This would argue that there is relatively little change in extinction as a function of radius.

We also evaluated the systematic
uncertainties in our results due to the measurement techniques, such as the
star masking, the particular choice of aperture radii, and local variations in
surface brightness due to the inherent distribution of individual stars around
\sgra. For example, we compared the variations in surface brightness   
distributions within the four separate image quadrants surrounding \sgra. 
The resulting distributions are shown in Figure 3.
We have used the uncertainty in the mean of these four measurements at a given
radius as an uncertainty in the azimuthally-averaged results shown in Figure 2. 

We also compared results using
different series of aperture radii, in order to determine if any bias is being
introduced by the large variation in aperture area from the smallest to the
largest apertures. Several different sets of aperture radii that used somewhat
more balanced numbers of pixels per aperture all yielded the same overall
profile shape, with the distinctive change in slope near
r$\sim$0.6--0.8\asec\ (as discussed in the following section). Finally, we also
computed the surface brightness using the median signal within each annulus and
compared it with the standard results that use the mean. As can be seen from
the comparison shown in Figure 4, both methods yield the
same slope in surface brightness distribution, to within the uncertainties of
the slope fits.

\section{Results and Discussion}

Plots of the surface brightness of stars as a function of distance from
\sgra\ for all three bands have been presented in Figure 2, both with and without
extinction corrections. Least-squares fits to the
surface brightness distributions are also superimposed. The slope of the
distribution in all three bands is significantly steeper at radii $>$ ~0.7\asec.
In each band we determined weighted fits using a broken, two-part power law.
The least-squares fits to the outer regions did not include the two
outermost data points (r$>$4\asec) due to a systematic drop in surface
brightness near that radius.
Table 1 shows the parameters of the individual fits in each wavelength band.
The fit parameters are quite consistent amongst the three bands. For
radii $>$ 0.7\asec, the slope is a relatively steep $\beta=-0.34\pm0.04$, 
where $\Sigma$  $\propto r^{\beta}$,  
and becomes flatter at smaller radii with $\beta=-0.13\pm0.04$.
We note that Sabha et al. (2010) have applied different methods of 
PSF subtraction to their AO data and have detected 
residual diffuse emission within 0.5$''$ of \sgra.  
These authors note a decrease in diffuse light away from \sgra\ with 
$\beta=-0.20\pm$0.05 over a range in projected distance of 0.03--0.2\asec, 
which is consistent with that of our measurements within r$\sim$0.7\asec. 

\subsection{Artifical Star Clusters}
In order to test whether the flattening of the surface brightness
distribution at small radii could be an artifact of our measurement
techniques, we applied the same measurements to artificial star clusters
and compared the results. We used tasks in the IRAF artificial data
(``artdata'') package to create the artificial clusters, using a NICMOS
camera 1 PSF generated by the STScI program ``Tiny Tim''. The task
``starlist'' was first used to generate a list of 2000 stars distributed
over an x/y coordinate space of the same extent as our NICMOS images,
with an $r^{-0.8}$ spatial distribution (corresponding to a space
density of $r^{-1.8}$) and a power-law luminosity distribution.
Ranges of artificial magnitudes were chosen to result in flux levels similar 
to those in the NICMOS images of \sgra. The ``mkobjects'' task was then used 
to produce an artificial image, using a NICMOS camera 1 PSF generated by the 
STScI program ``Tiny Tim'' to create suitable profiles at each of the
positions in the star list, along with appropriate levels of Poisson and
read noise. Ten separate artificial clusters were generated in this way, 
using different seeds for the random number generators that produce the
star catalog. 

Two images were produced for each of the ten artificial
clusters, which allowed us to assess the effectiveness of the masking
technique that we applied to the bright stars in the NICMOS images. We
generated one image of each cluster that contained all the bright stars in
the catalog produced by ``starlist'' and then applied masking to the bright 
stars over the same range
of magnitudes as that in the NICMOS images. We also generated an image for
each cluster that excluded the set of bright stars from the catalog, so
that the bright stars did not appear in the images. We then measured the 
surface brightness distributions in each of the cluster images using the 
same technique as that applied to the NICMOS data. 
Figure 5 shows the results.

The upper set of points in Figure 5 shows the average surface brightness
distribution for the ten artificial clusters that did not contain any of
the bright stars. Similarly, the lower set of points corresponds to the
data for the clusters in which the bright stars were masked out. The two
sets of points have been offset from one another along the y-axis of the plot
for the sake of clarity. The error bars indicate the standard deviation
of the values amongst the ten simulations at each radius. The spread in
data values is naturally somewhat larger for the clusters that had
masking applied, due to the smaller number of data points used at each
radius.

There are two important points to notice from these results. First, there is
very little difference between the results derived from the masked images
and the images that did not include the bright stars. The slopes of the 
least-squares fits are identical to within the fit uncertainties. This
indicates that the masking technique that was applied to the
NICMOS images is effective at removing the contributions of the bright
stars near \sgra. Second, the surface brightness profiles of
both versions of the artificial clusters are well-fit by a single power law,
with very little deviation from the linear fit.
In contrast to the measurements made from the NICMOS images of \sgra,
there is no indication of a different slope for radii $<$ 0.7\asec. 
Hence the flattening of the \sgra\ profile at small radii does not appear to 
be an artifact of the data or the measurement techniques.

\begin{deluxetable}{lll}
\tablewidth{0pt}
\tablecaption{Parameters of the Fit in Three Bands}
\tablehead{
\colhead{Band ($\mu$m)}&
\colhead{Radii ($''$)}&
\colhead{Slope ($\beta\pm\sigma$)}
}
\startdata
 1.70             & $<0.7$  & -0.13$\pm0.04$ \\
  -            & $0.7-4.7$  & -0.34$\pm0.04$ \\
 1.45             & $<0.7$  & -0.14$\pm0.04$ \\
 -             & $0.7-4.7$  & -0.39$\pm0.04$ \\
 1.90             & $<0.7$  & -0.15$\pm0.07$ \\
 -             & $0.7-4.7$  & -0.31$\pm0.04$ \\
\enddata
\label{tab:fitparams}
\end{deluxetable}

\subsection{ Flattening of the Surface Brightness Distribution}
The flattening of the slope in the two-part fit is consistent with AO 
measurements (Sch\"odel et al. 2007) in that the slope becomes steeper with 
increasing radii, with $\beta$=--0.13 for $r < 0.7$\asec\ and $\beta$=--0.34 
for $r = 0.7-4.7$\asec. The distribution of total light within 0.7\asec\ has
a similar slope, because the disks of massive stars are located beyond 0.8$''$.
AO measurements show a flattening of the evolved faint stars after having
identified and removed the contributions of B stars. To make both measurements consistent 
with each other, the NICMOS data must be contaminated by faint members of the 
S cluster within 1$''$ of \sgra. Alternatively, because we have sampled the 
diffuse stellar emission within the inner 1$''$ better than stellar counts 
from AO measurements, the NICMOS data includes the background light that may 
have not been detected or included in the AO results. 
To understand the origin of this discrepancy, we have converted the observed 
flux of 7.72 mJy from the region within 1\asec\ of \sgra\ and estimated 
an equivalent H-band magnitude, which is the band closest to our 1.7\mic\ band, of 12.8. 
Using the observed $H-K$ color of ~2 for a population of 
evolved stellar sources in the Galactic center region (Maness et al. 2007), 
we expect K=10.8 mag within 1\asec\ of \sgra. 
This is equivalent to $\sim$76 stars of K=15.5 mag in the inner 1\asec. 
Number counts of stars within 1\asec\ clearly show a smaller number of 
K=15.5 stars than this (e.g., Bartko et al. 2010; Buchholz et al. 2009), 
suggesting that this light comes from dwarf stars rather
than the giants that would be detectable in the AO
observations (see \S 3.3). The equivalent number of stars also 
exceeds the extrapolations of the luminosity 
functions  based on
the number counts, consistent with the excess being the integrated
light from dwarves.


The measurements presented here also confirm that the sudden drop in the
radial profile of starlight near r=0.8\asec\ reported by Scoville et al. (2003)
does not actually exist.  Scoville et al. (2003)  attempted
to remove the contribution of massive stars by discarding the signal from the
brightest 20\%\ of the pixels in each photometric annulus around \sgra.
This, however, produces a variable rejection threshold as a function of radius,
depending on the maximum brightness of stars included in each annulus.
Figure 6, which shows a comparison of the surface brightness distribution with
and without rejection of the bright stars, shows the complexity and dominance 
of the light from bright stars outside 1\asec\ radius.  The bright stars (shown
by the blue line in Figure 6) contribute a large amount of signal at
$r > 0.8$\asec, which produces an obvious bump in the distribution. The dip that was 
reported by Scoville et al. (2003) inside of r=0.8\asec\ is in fact
simply due to this peak in the number of bright stars just outside that radius,
which was not completely removed from their measurements.
If the bright stars are completely masked, the dip in the surface brightness 
near 0.8\asec\ disappears, as shown in Figure 6.




\subsection{ Model Fitting}
We have also fit our surface brightness distributions using a parametrized form of 
the three-dimensional luminosity density profile and projecting it onto the plane 
of the sky. We follow Merritt (2010) and adopt a spherically-symmetric 
luminosity-density profile for the stellar cluster, using
\begin{equation}
	\rho_L(r) = \rho_0 \left(\frac{r}{r_0}\right)^{-\gamma_i} 
    \left[1+\left(\frac{r}{r_0}\right)^{\alpha}\right]^{(\gamma_i-\gamma)/\alpha} \,,
	\label{eq:rho}
\end{equation}
which behaves as $r^{-\gamma_i}$ for $r\ll r_0$, and $r^{-\gamma}$ for $r\gg
r_0$, with $\alpha$ controlling the sharpness of the transition from the
outer to inner power-law around $r_0$.  We also follow Merritt (2010) in fixing
$\alpha=4$ and, because our data do not extend to very large radius, impose
$\gamma=1.8$ to follow previous determinations at large $r$.
We integrate equation 1  along the line of sight to compute the
surface brightness profile and adjust the parameters $n_0$,
$r_0$, and $\gamma_i$, for each band, to minimize $\chi^2$.

\begin{table}[tbp]
	\centering
	\caption{Fits to surface brightness profile}\vspace*{12pt}
	\begin{tabular}{ccccccrr}
\hline
\hline
	fit & $\rho_0$ 1.45$\mu$m       & $\rho_0$ 1.70$\mu$m      & $\rho_0$ 1.90$\mu$m     & $r_0$  & $\gamma_i$ & $\chi^2/\textrm{d.f.}$\\[3pt]
	    & (mJy\,arcsec$^{-3}$) & (mJy\,arcsec$^{-3}$)  & (mJy\,arcsec$^{-3}$) & (arcsec) & & \\[3pt]
\hline
	1.45 $\mu$m only & 15.03 &     --- &     --- &  3.37 & 0.94 &  0.63 \\[3pt]
	1.70 $\mu$m only &     --- & 11.19 &     --- &  3.73 & 0.87 &  0.59 \\[3pt]
	1.90 $\mu$m only &     --- &     --- &  8.40 &  4.36 & 0.92 &  0.28 \\[3pt]
	    simultaneous & 13.31 & 11.22 & 10.75 &  3.69 & 0.89 &  0.55 \\[3pt]
	  $\gamma_i=0.5$ & 28.10 & 23.97 & 22.66 & 2.36 & 0.50 &  0.94 \\[3pt]
	 $\gamma_i=1.25$ &  0.647 &  0.536 &  0.524 & 27.4 & 1.25 &  1.28 \\
         fixed $r_0$           & 6.23   &  5.24  &  5.04  & 6.00    & 1.06 &  0.73 \\
\hline
	\end{tabular}
	\label{tab:fits}
\end{table}


Figure 7 and Table 2 compare individual profiles from NICMOS data in the 1.45, 1.7 and 1.9$\mu$m filters with 
simultaneous fits across all three bands, as shown in blue. For the individual fits, the values of $r_0$ and 
$\gamma_i$ increase with wavelength, which means that there is very little difference between the simultaneous fits 
and the 1.7$\mu$m fit, because the differences in the shapes at 1.45 and 1.9$\mu$m relative to that at 1.7$\mu$m 
tend to cancel. The reduced $\chi^2$ drops in moving between the separate fits for the 1.7, 1.45, and 1.9$\mu$m 
data simply because of the decrease in S/N between the bands. Overall, the values of reduced $\chi^2$ that we 
obtain are small, suggesting that the error bars may be overestimated. Therefore, we cannot use the magnitude of 
$\chi^2$ to reliably determine confidence limits for the model parameters. This limitation aside, the best fit to 
the data suggests that the stellar luminosity density profile flattens as $\sim r^{-0.9}$ within 3--4\asec\ of 
\sgra. We do not find evidence for a decline in stellar density (i.e. negative values of $\gamma_i$) in the inner 
arcsecond, in contrast to the results of Do  et al. (2009; see also Merritt 2010). The best-fit for an 
$r^{-0.5}$ inner profile, as shown with red short-dashed line, is only marginally consistent with the observed 
surface density profile.

Our values of $r_0$ are not very well determined as our data extend only to 
5$''$ from
Sgr A*.  To assess this uncertainty we also
consider a fit with $r_0$ fixed at 6$''$ (dashed blue curve in Fig.~7), consistent 
with
the profiles found by Sch\"odel et al. (2007)
and Bartko et al. (2010).  This yields $\gamma_i=1.06$ and a slightly larger
value of the $\chi^2$ per degree of freedom, and so
we conclude that  $\gamma_i$ may be larger than the 0.89 we obtain when $r_0$ is
allowed to be adjusted. 

Converting our measured diffuse light profile to a stellar mass profile 
is not straightforward.  The intrinsic colors inferred 
from the ratios of the $\rho_0$ values derived from simultaneously fitting the profiles in the three NICMOS bands are consistent 
with either giant stars with spectral types of G4\,III -- K0\,III, or K0--K2 dwarfs.  The former would be bright enough (extinction-corrected 
H$\la 13$) to be detected as point sources, so it is reasonable to assume that the light is dominated by K0 dwarfs, which 
would have an extinction-corrected $H\approx 18.7$ (corresponding to an F170M flux of about $4\ee -5 $\,Jy)  and mass $\approx 
0.8$M$_\odot$ (e.g.\ Binney \& Merrifield 2000).  The corresponding conversion factor between F170M flux and stellar mass is 
then $2.05\ee 4 \,\mathrm{M}_\odot \mathrm{Jy}^{-1}$, and the luminosity density at 1.7\micron\ 
implies that the mass enclosed 
within $r\la 0.1$\,pc is $\approx 3.2\ee 4 $\,M$_\odot (r/0.1\,\mathrm{pc})^{2.1}$.  
This is in reasonable agreement with 
Sch\"odel et al (2009), who used the diffuse light at 10\asec\ in their $K_s$ band observations, the radial profile derived from 
counts of late-type stars, and a solar K-band luminosity-to-mass ratio to find $M(r)\approx 1.9\ee 4 \, \mathrm{M}_\odot 
(r/0.1\,\mathrm{pc})^{1.8}$ for $r\la 0.8\,$pc.  Adjusting Sch\"odel et al's $A_K=3.3$\,mag to our adopted $2.42$\,mag decreases 
this mass estimate by a factor of 0.44, yielding $0.85 \ee 4 \, \mathrm{M}_\odot $ enclosed within 0.1\,pc.  For comparison, if 
we also assume that the diffuse light was dominated by G2 dwarfs, (extinction-corrected $H\approx 17.9$), our estimate of the 
stellar mass enclosed within 0.1\,pc is $2.3\ee 4 $\,M$_\odot$. 
These masses are all consistent with the analysis of proper 
motions measurements, which indicate a similar mass enclosed near \sgra\ (Sch\"odel et al. 2009). 
Morris (1993) claims the possibility of several 10$^5$ \msol of stellar
mass black holes in the innermost tenths of parsec of Sgr A*.  Our
analysis does not
support the evidence for such a  population of stellar remnants
close to Sgr A*.

Our measurement of the surface brightness within a 10\asec\ field centered on \sgra\ is consistent with an 
$r^{-0.9}$ luminosity density profile in the innermost 0.1\,pc, similar to the $r^{-1.12}$ power law inferred by 
Sch\"odel et al.\ (2009) from counts of late-type stars.  For such a flat inner profile, the surface brightness is 
insensitive to reductions in the luminosity density within 0.01\,pc of Sgr A*, because the surface brightness is 
dominated by the luminosity density at larger radii.  
Nevertheless, our analysis finds no evidence of reduction in the diffuse   
light distribution within 0.1 pc. In contrast, Buchholz et al. (2009)
found a reduction
in the number  counts of starts brighter than 15.5 K magnitude in the
inner 0.1 parsec.
These are marginally consistent with an $r^{-1}$ profile (Merritt 2010), but may suggest preferential 
removal of giants.

One mechanism that can destroy the $r^{-1.75}$ cusp expected from a relaxed
system of stars near a supermassive black hole (Bahcall \& Wolf 1976) is
mass segregation, with black holes tending to settle towards the center
while boosting stars to larger radii (e.g.\ Preto \& Amaro-Seoane
2010).  The resulting inner power law, $\gamma_i\approx-1.25$, is still too
steep to be consistent with our data, as shown by the corresponding
surface-density profiles (long-dashed red curves) in Figure 7.  An alternative
is that some process has removed the lowest energy stars near the center,
which results in a shallower power law in the core, down to a limiting
value (for isotropic velocity distributions) of $\gamma_i= 0.5$, that can
survive for several Gyr (Merritt 2010).  This limiting value of $\gamma_i$
appears to be marginally consistent with our surface brightness profile, 
and larger values are more consistent with our measured profile.
Another process that could slow down the two-body relaxation time scale is
the presence of extended clouds from which the massive stars formed during the  history 
of star formation in the nuclear star cluster. Future numerical simulations
should include the effects  of star forming gas clouds 
in the evolution of stellar clusters that surround   massive black holes.



Acknowledgments: We thank the referee for many excellent and useful comments.
 This work is partially supported by grants 
AST-0807400
from the National Science Foundation and DP0986386 from the Australian Research Council. 



\begin{figure}
\center
\includegraphics[scale=0.55,clip, trim=12mm 15mm 25mm 25mm]{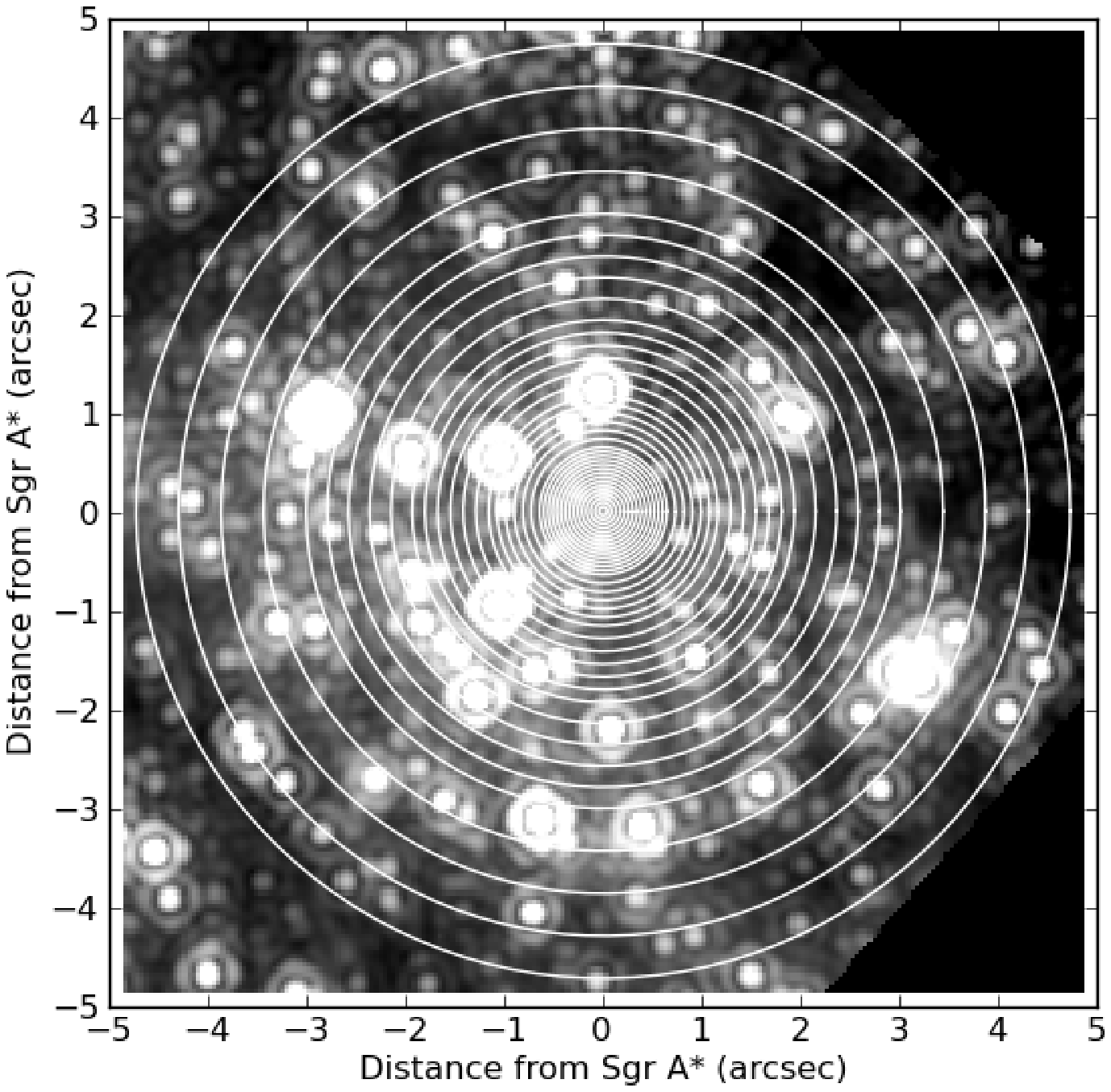}
\includegraphics[scale=0.55,clip, trim=12mm 15mm 25mm 25mm]{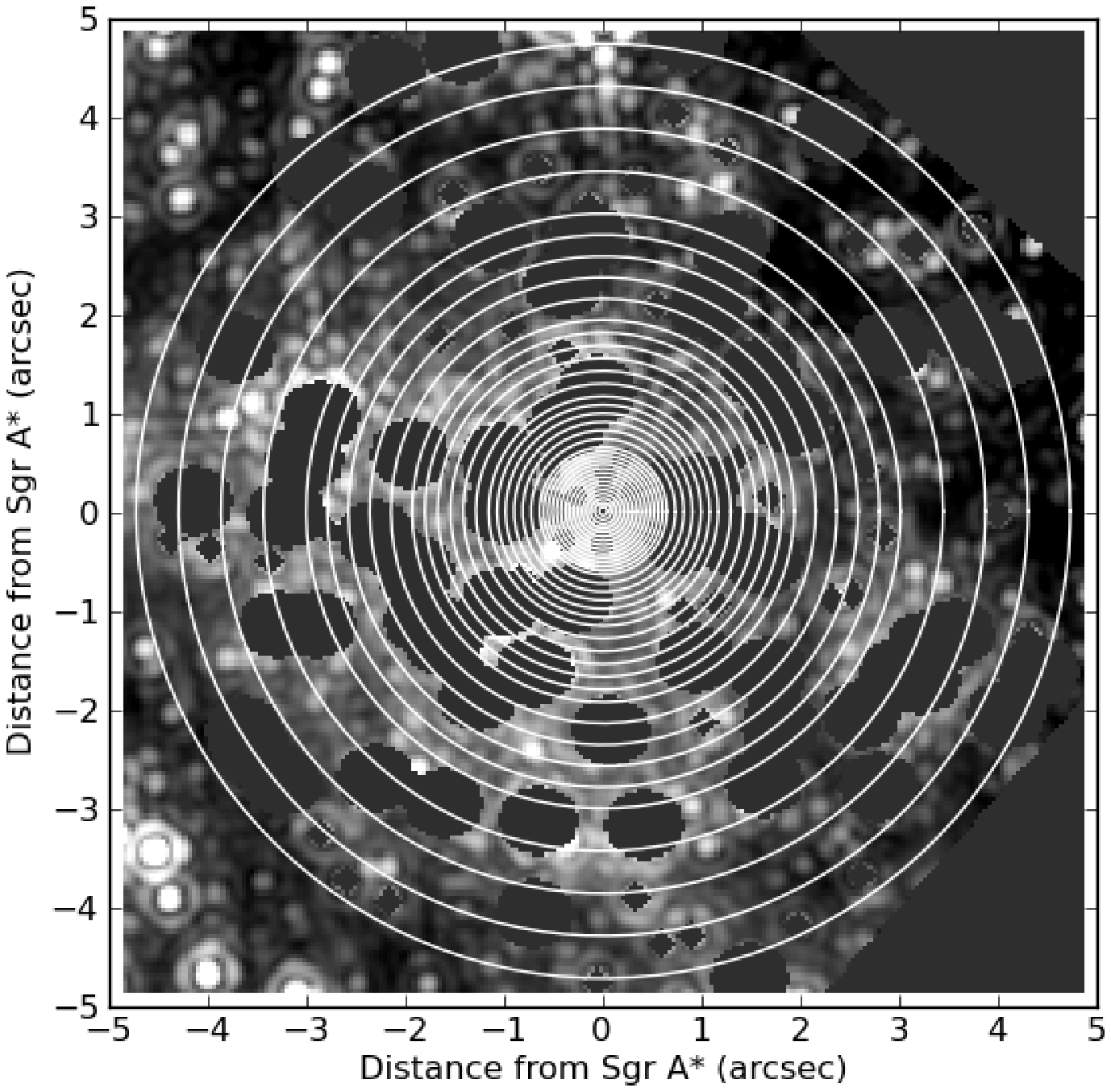}\
\includegraphics[scale=0.55,clip,trim=10mm 15mm 22mm 25mm]{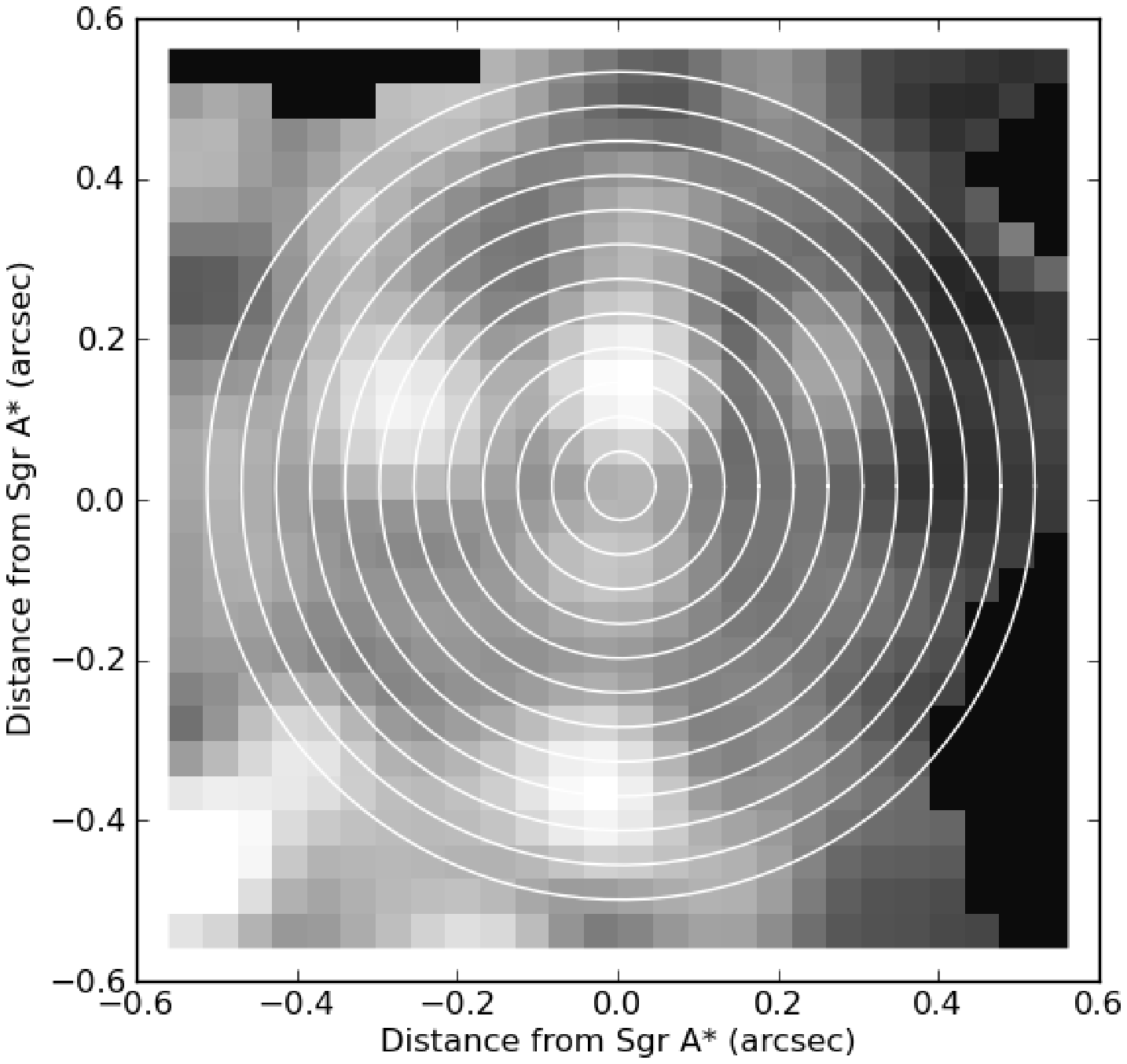}
\includegraphics[scale=0.55,clip,trim=10mm 15mm 22mm 25mm]{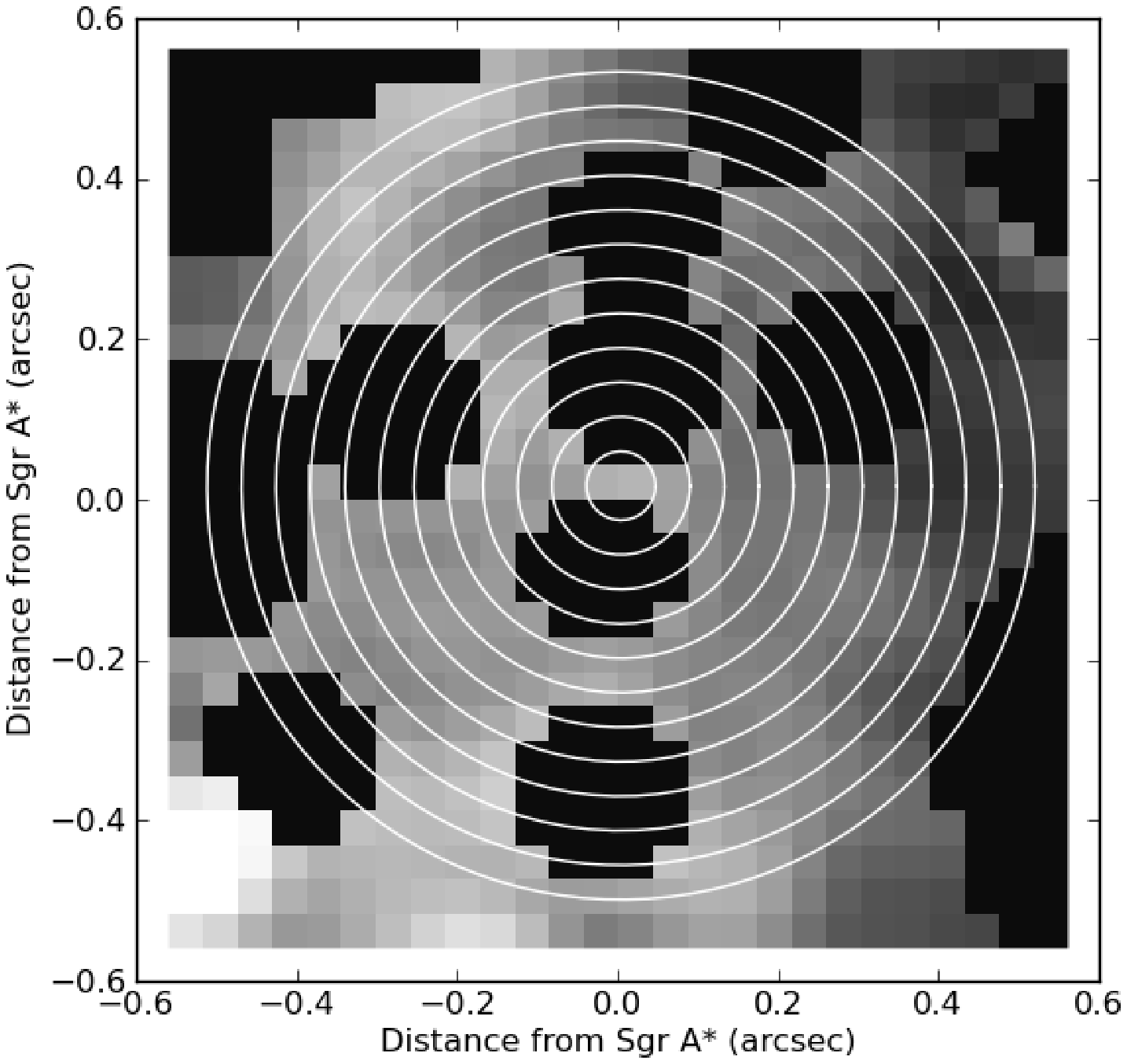}
\caption{
{\it (a) Top Left}
A greyscale image of the inner 10$''$ of the Galactic center centered on 
the position of \sgra\ at the wavelength of 1.7$\mu$m. 
A series of photometry apertures are overlaid on the image in order
to determine the surface brightness as a function of radius. 
Each pixel in the image  is 0.043$''$. 
{\it (b) Top Right} Similar to (a) except  that the bright stars have been masked. 
{\it (c) Bottom Left}
A close-up view of the inner 1$''$ of the Galactic center 
is shown. The brightest  stars are members of the S cluster. 
The overlaid photometry apertures are also shown.  
{\it (d) Bottom Right} Similar to (c) except that the bright S stars, including S0-2,
have been masked. The stars S34+S35 at the southwest corner are considered to be evolved
and have not been masked.
}
\end{figure}

\begin{figure}
\center
\includegraphics[scale=0.5,clip,angle=0,trim=15mm 0mm 25mm 0mm]{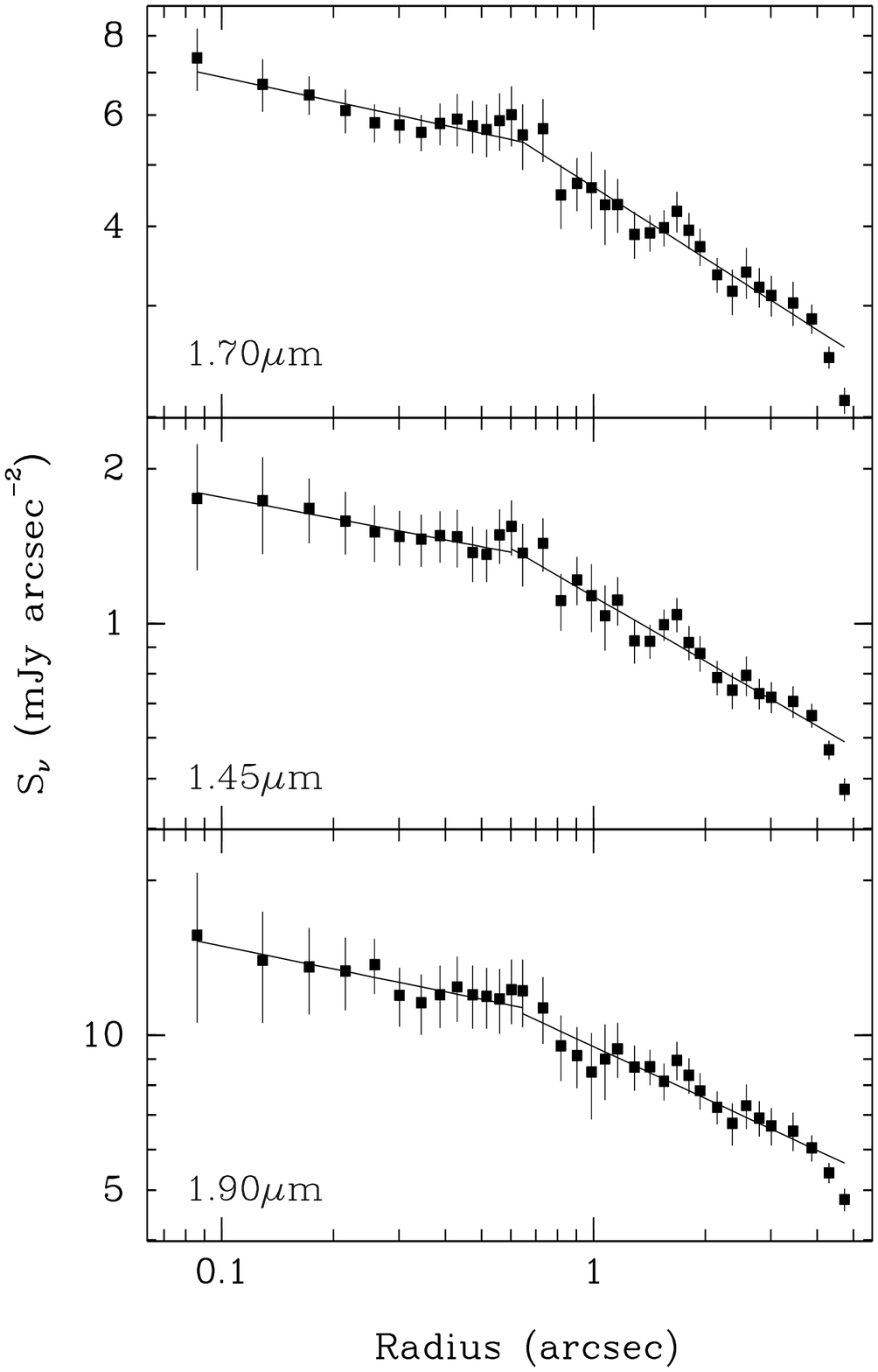}
\includegraphics[scale=0.5,clip,angle=0,trim=10mm 0mm 25mm 0mm]{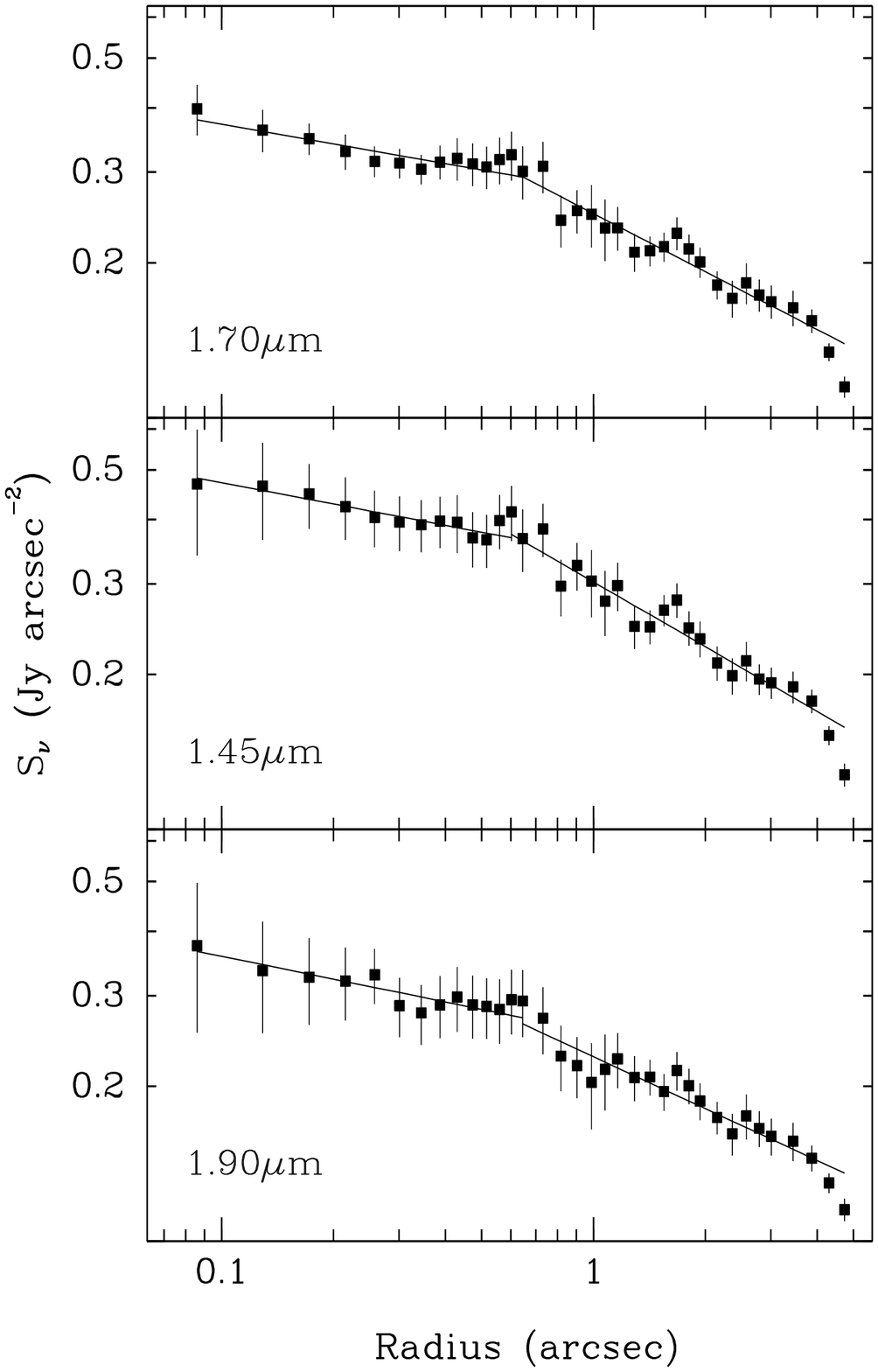}
\caption{
{\it (a) Left}
The observed surface brightness as a function of radius in the 3 NICMOS
bands. A two-part fit is superimposed  on the radial profile with parameters 
as shown in Table 1. 
{\it (b) Right} Same as (a) except that the fluxes are extinction corrected
using $A_{1.45\mu m}=6.07$, $A_{1.7\mu m}=4.34$, and $A_{1.9\mu m}=3.46$.  
}
\end{figure}

\begin{figure}
\center
\includegraphics[scale=0.6,angle=-90]{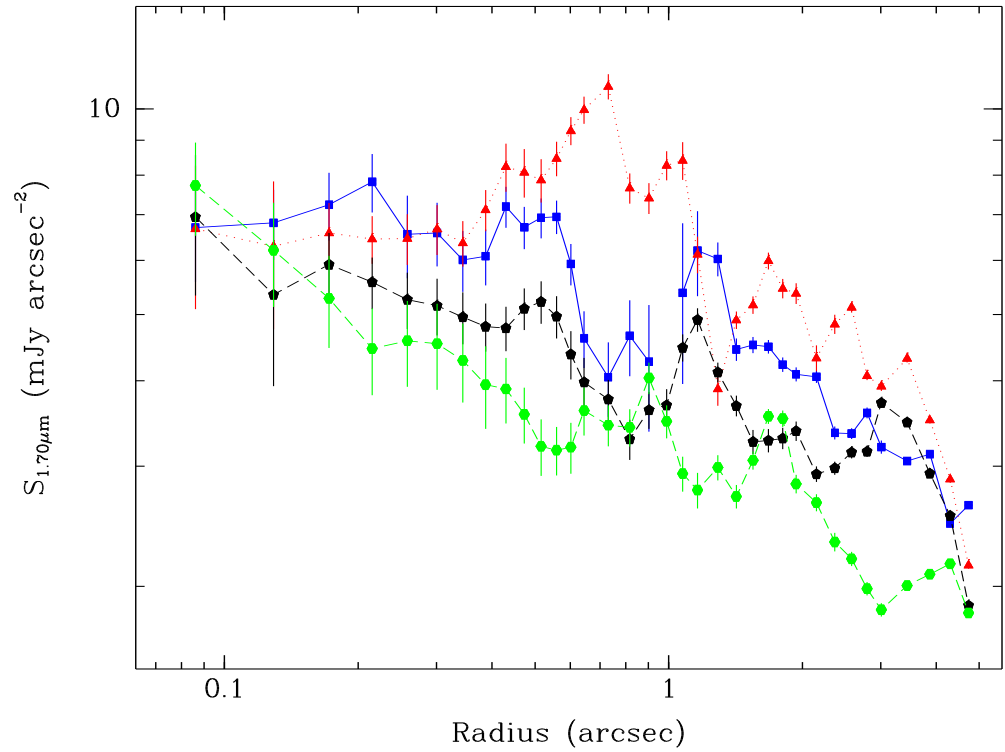}\\
\caption{
The observed surface brightness as a function of radius at 1.7$\mu m$
measured separately for each of the four image quadrants surrounding
\sgra. The error bars shown in Figure 2 are the uncertainty in the mean
at a given radius of these data.
The colors for 
quadrants  1, 2, 3 and 4 are blue, red, black and green, respectively. 
The first quadrant is  
in the upper-left corner of the image (NE on the sky) 
and rotate around counterclockwise. 
}
\end{figure}

\begin{figure}
\center
\includegraphics[scale=0.6,angle=-90]{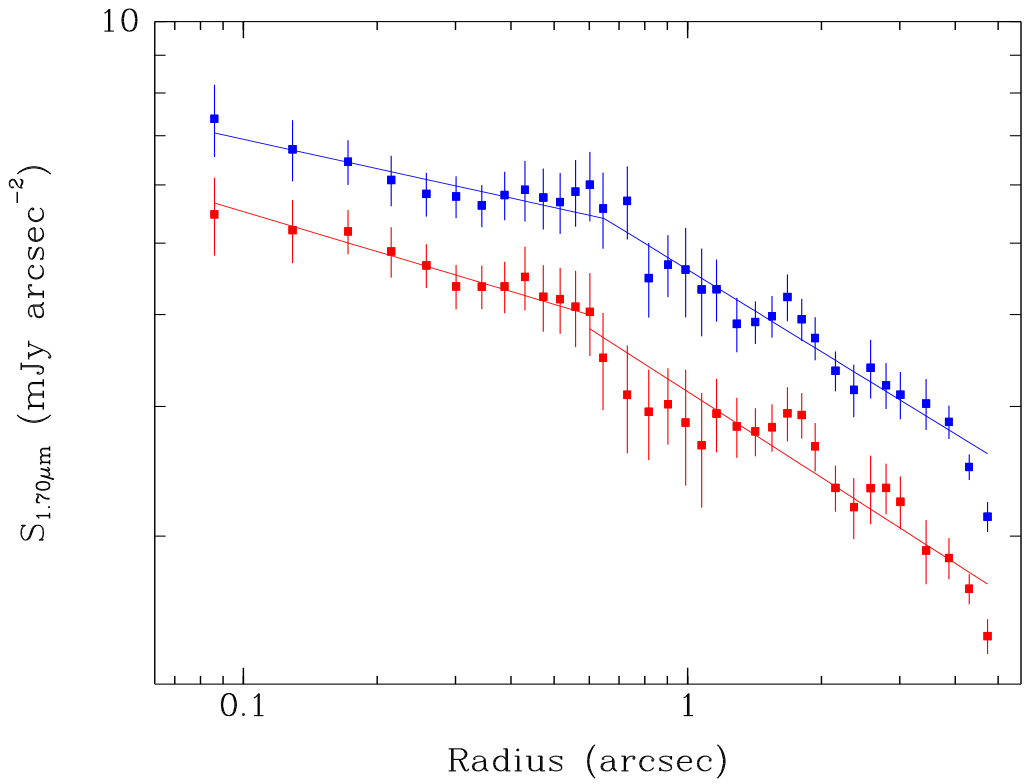}\\
\caption{
The observed surface brightness as a function of radius at 1.7$\mu m$
using the mean (upper) and median (lower) of the unmasked pixel values
within each photometric annulus. The two curves has been offset vertically
for clarity. The slopes for the two methods are identical to
within the uncertainties.
}
\end{figure}

\begin{figure}
\center
\includegraphics[scale=0.6,angle=-90]{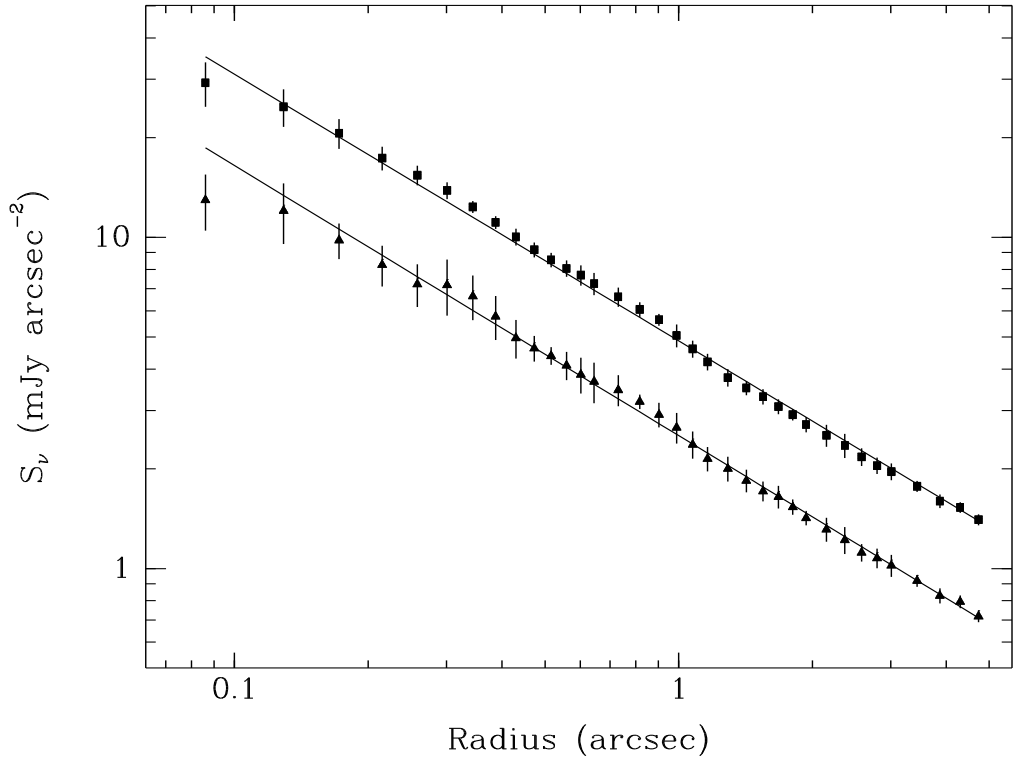}\\
\caption{
Averaged profiles and fits for the 10 artificial clusters. The upper line, 
with square markers, is the data for the clusters that did not have the bright 
stars included in the images. The lower line, with triangular markers, is for 
the clusters that had the brights stars masked. The two sets of data have been
artificially separated along the y-axis for clarity.
Error bars are the standard deviation amongst the 10 clusters at each radius.}
\end{figure}

\begin{figure}
\center
\includegraphics[scale=0.6,angle=-90]{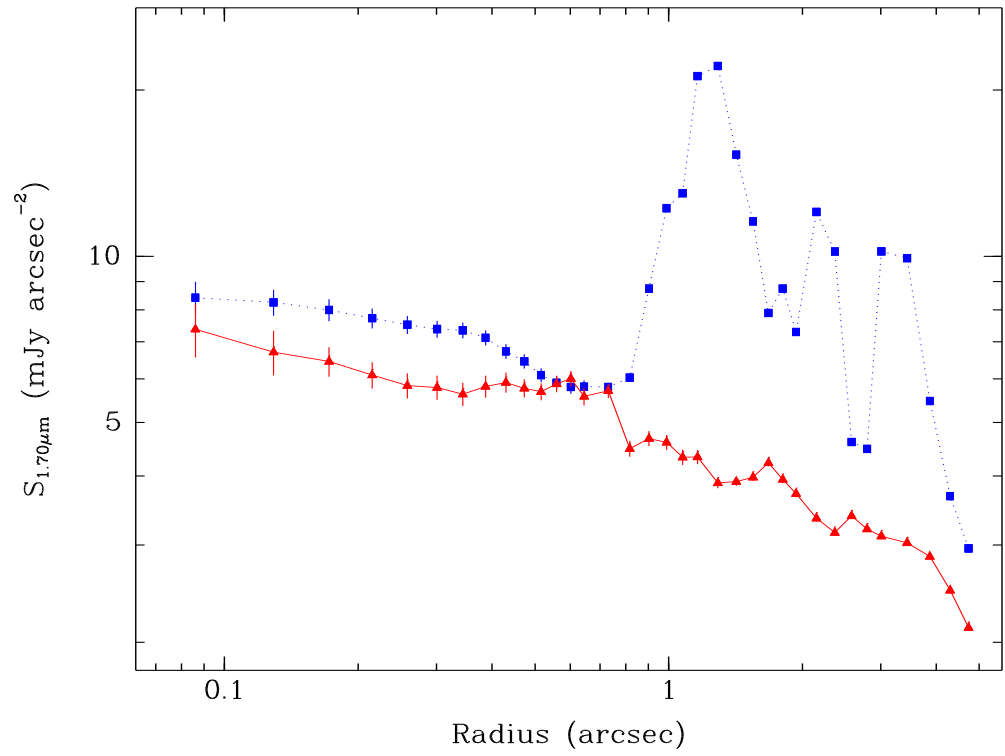}\\
\caption{
A comparison of the 1.7\mic\ surface brightness distributions measured
with (red) and without (blue) masking of the bright, young stars in the NICMOS image. 
}
\end{figure}

\begin{figure}
\center
\includegraphics[scale=0.8,angle=0]{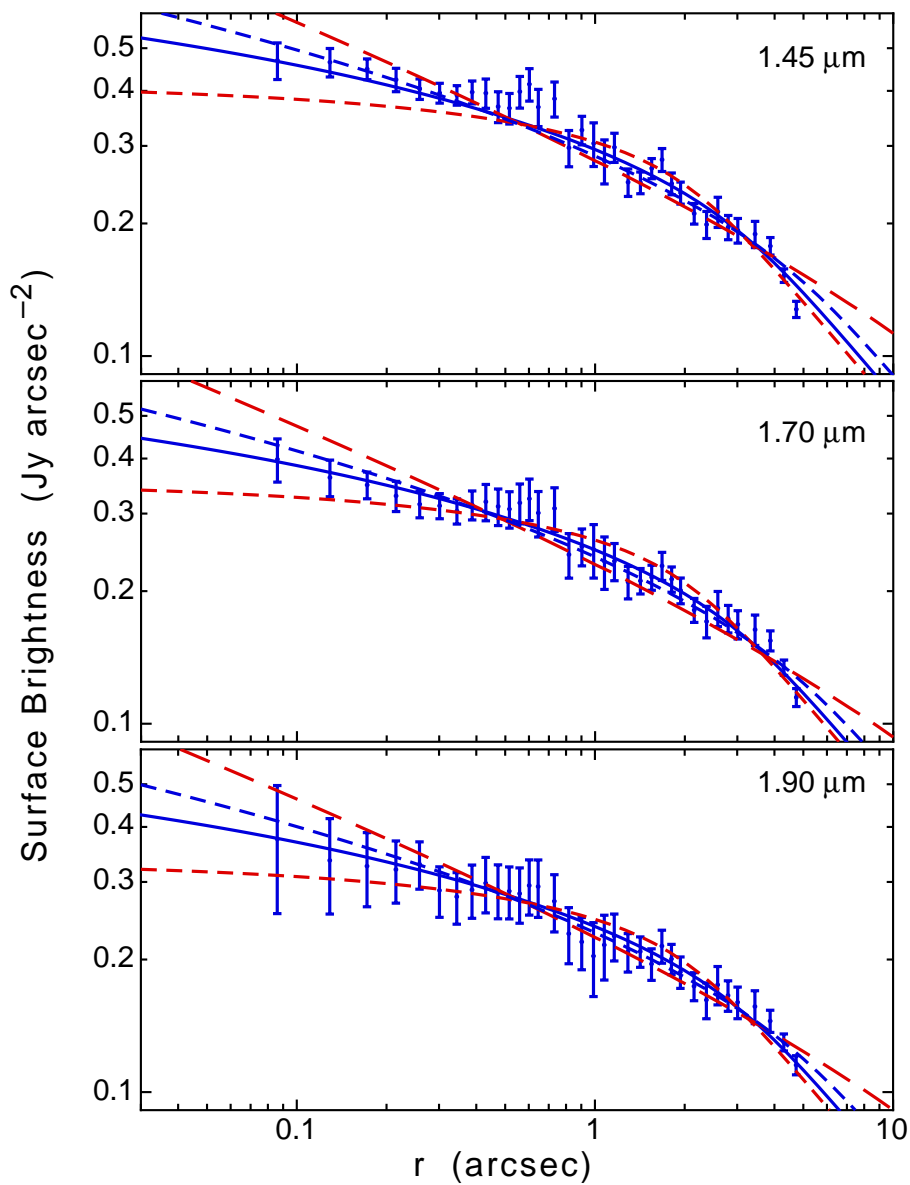}
\caption{
Sample fits to the HST data.  \textit{Blue solid curves} show the best
                simultaneous fit across all 3
                bands.  \textit{Red short-dashed} and \textit{long-dashed} 
curves show the
                best simultaneous fits obtained when $\gamma_i$ is fixed at 
               0.5 or 1.25,  respectively (see text). 
               \textit{Blue short-dashed} curve shows the simultaneous fit when 
$r_0$ is fixed at 6" 
(see text)
}
\end{figure}

\end{document}